\documentclass[%
 reprint,
 amsmath,amssymb,
 aps,
]{revtex4-1}
\usepackage{setspace}
\usepackage{graphicx}
\usepackage{dcolumn}
\usepackage{bm}

\begin{document}

\preprint{APS/123-QED}

\title{Ab initio no-core properties of $^{7}$Li and $^{7}$Be with JISP16 and NNLO$_{\rm{opt}}$ interactions}%
\author{Taihua Heng$^{1,2}$, James P. Vary$^{2}$ and Pieter Maris$^{2}$}
 \altaffiliation{hength@ahu.edu.cn, hength@iastate.edu,\\
  jvary@iastate.edu, pmaris@iastate.edu}
\affiliation{%
\small $^{1}$School of Physics and Material Science, Anhui University, Hefei, China, 230601\\
\small $^{2}$Department of Physics and Astronomy, Iowa State University, Ames, IA, United States, 50011
}%

\date{\today}
\begin{abstract}
We investigate the properties of $^{7}$Li with the JISP16 and chiral NNLO$_{\rm{opt}}$ nucleon-nucleon interactions
and $^{7}$Be with the JISP16 interaction in the $ab$ $initio$ no-core full configuration (NCFC) approach.
We calculate selected observables that include energy spectra, point proton root-mean-square radii, electromagnetic moments and transitions.
We compare our results with experimental results, where available, as well as with results obtained using nucleon-nucleon
plus three-nucleon interactions. We obtain reasonable agreement between theory and experiment for low-lying states
that are dominated by $p$ -shell configurations.
\\
\begin{description}

\item[PACS numbers]
21.60.De, 27.20.+n, 21.10.Ky

\end{description}
\end{abstract}

\maketitle

\section{\label{sec:level1}Introduction}

~~~An outstanding problem in nuclear physics is to study the properties of atomic nuclei based on realistic interactions among the protons and neutrons. The rapid development of $ab$ $initio$ methods for solving finite nuclei provides a theoretical foundation that can be numerically solved to high precision using realistic nucleon-nucleon ($NN$) and three-nucleon ($NNN$) interactions.
A number of meson-exchange potentials, sometimes combined with phenomenological terms to achieve high accuracy in fitting $NN$ data (CD-Bonn \cite{MacPRC63}, Nijmegen \cite{StoPRC49}, Argonne \cite{WirPRC51}), have been developed that should be used together with modern $NNN$ forces (Urbana \cite{CarNPA401}\cite{PudPRC56}, Illinois \cite{PiePRC64}, Tucson-Melbourne \cite{CooNPA317}\cite{FriPRC59}\cite{HubFBS30}) to describe properties of many-body nuclear systems. A very important step in the theory of inter-nucleon interactions is the emergence of realistic $NN$ and $NNN$ interactions tied to quantum chromodynamics (QCD) via Chiral Perturbation Theory (CPT) \cite{BedPRL82}\cite{EpePRC66}\cite{EntPLB524PRC68}.\\

In addition, recent advances in the utilization of high-performance computing systems offer an opportunity for $ab$ $initio$ approaches to be at the forefront of nuclear structure explorations. Various microscopic many-body methods have been developed for no-core treatments of nuclei, including No-Core Shell Model (NCSM) \cite{NavPRL84}\cite{NavPRC62}\cite{JPV348}, the Green's Function Monte Carlo (GFMC) approach \cite{PiePRC70}\cite{PerPRC76}\cite{MarPRC78}, the Coupled-Cluster (CC) method \cite{HagPRL103}\cite{HagPRC82}, etc. The name NCSM is due to the fact that all the nucleons of the nucleus are active and treated on an equal footing and thus no inert core is assumed. The NCSM method was introduced as a finite matrix truncation of the infinite matrix problem with a renormalized Hamiltonian specific to that truncation.\\

 Here we adopt the $ab$ $initio$ No-Core Full Configuration (NCFC) approach \cite{MarPRC79}\cite{JPV342}\cite{BogNPA801} which is an extension of the NCSM approach. In
 the NCFC approach, the direct solution of the nuclear many-body problem is obtained by diagonalization in a sufficiently large basis space that converged binding energies are obtained-either directly or by simple extrapolation. \\

 We select a traditional harmonic oscillator (HO) basis so there are two basis space parameters, the HO energy $\hbar\Omega$ and the many-body basis space cutoff $N_{\rm{max}}$. $N_{\rm{max}}$ is defined as the maximum number of total oscillator quanta allowed in the many-body basis space above the minimum for that nucleus. We obtain convergence in this two-dimensional parameter space ($\hbar\Omega, N_{\rm{max}}$), where convergence is defined as independence of both parameters within estimated uncertainties.\\

In the present work, we investigate the properties of $^{7}$Li and $^{7}$Be in the $ab$ $initio$ NCFC approach with the JISP16 \cite{ShiPRC70}\cite{ShiPLB621}\cite{ShiPLB644} and chiral NNLO$_{\rm{opt}}$ \cite{EksPRL110} interactions. The JISP16 $NN$ interaction, proposed in Ref. \cite{ShiPLB644}, is constructed in the J-matrix inverse scattering approach \cite{ShiPRC70}\cite{Alhbook}. It is known to provide an excellent description of $np$ scattering data with $\chi^{2}/datum=1$ \cite{ArnPRC76}. The interaction was fitted in Ref. \cite{ShiPLB644} by means of phase-equivalent transformations to a few binding energies of nuclei with $A\leq 16$, and it provides a good description of bindings and spectra of light nuclei without referring to three-nucleon forces \cite{ShiPLB644}\cite{ShiPAN71}\cite{MarPRC81}\cite{Shibook}\cite{JPV359}.\\

Chiral effective field theory (EFT) is a promising theoretical approach to obtain a quantitative description of the nuclear force from first principles \cite{MacPR503}. Interactions from chiral EFT employ symmetries and the pattern of spontaneous symmetry breaking of QCD \cite{MacPR503}\cite{EpeRMP81}. Moreover, the interaction is parametrized in terms of low-energy constants (LECs) that are determined by fitting experimental data. Our adopted NNLO$_{\rm{opt}}$ interaction at next-to-next-to-leading order (NNLO) is constructed by the optimization tool POUNDerS (Practical Optimization Using No Derivatives for Squares) in the phase-shift analysis \cite{KorPRC82}\cite{MunANL2012}. The optimization of the low-energy constants in the $NN$ sector at NNLO yields a $\chi^{2}$/datum of about one for laboratory scattering energies below 125 MeV. The NNLO$_{\rm{opt}}$ $NN$ interaction is also fitted to provide very good agreement with binding energies and radii for A = 3 and 4 nuclei. Some key aspects of nuclear structure, such as excitation spectra, the position of the neutron drip line in oxygen, shell-closures in calcium, and the neutron matter equation of state at subsaturation densities, are reproduced by the NNLO$_{\rm{opt}}$ interaction without the addition of three-nucleon forces.

\section{$ab$ $initio$ no-core full configuration (NCFC) approach}

~~~We give a brief review of the NCFC method and, for more details, we refer the reader to Refs. \cite{MarPRC79}\cite{JPV342}\cite{JPV359}\cite{Shibook}. The Hamiltonian for the A-body system in relative coordinates is
\begin{equation}
H_{A}=\frac{1}{A}\sum_{i<j}\frac{(\vec{p}_{i}-\vec{p}_{j})^{2}}{2m}
+\sum_{i<j}V_{NN}(\vec{r}_{i}-\vec{r}_{j})+\sum_{i<j}V_{C}(\vec{r}_{i}-\vec{r}_{j}),
\end{equation}
where $V_{NN}(V_C)$ is the two-nucleon (Coulomb) interaction. Our expression for $H_A$ is somewhat schematic since the $NN$ interaction may, in general, be non-local. Furthermore, in the present work, we neglect $NNN$ interactions. We solve the corresponding Schr\"{o}dinger equation
\begin{equation}
H_A\Psi(\vec{r}_1,\vec{r}_2,...,\vec{r}_A)=E\Psi(\vec{r}_1,\vec{r}_2,....\vec{r}_A)
\end{equation}
with numerical techniques formulated within a basis expansion (configuration interaction) approach. In the NCFC, the wave function $\Psi(\vec{r}_1,\vec{r}_2,....\vec{r}_A)$ is a superposition of Slater determinants  $\Phi_i$ of single-particle HO states.
\begin{equation}
\Psi(\vec{r}_1,\vec{r}_2,....\vec{r}_A)=\sum_{i}c_{i}\Phi_i(\vec{r}_1,\vec{r}_2,....\vec{r}_A)
\end{equation}
\\

For practical calculations, a finite basis space is specified by the $N_{\rm{max}}$ truncation. We aim for results that are convergent as determined by their independence of the parameters $\hbar\Omega$ and $N_{\rm{max}}$ as $N_{\rm{max}}$ is increased. We present results for even values of $N_{\rm{max}}$ that correspond to states with the same parity as the lowest HO configuration (the $N_{\rm{max}}=0$ configuration) and are called the "natural" parity states.\\

 We also need to address the issue of center of mass (COM) motion. In the above discussion, the Hamiltonian is in the relative coordinate and not in the single-particle coordinates where the HO states are specified. The methods of solution, including the method to constrain the COM motion, are not described in this paper, because these details have been explained in Refs. \cite{MarPRC79}\cite{JPV342}\cite{JPV359}\cite{Shibook}. We obtain our results with the MFDn code, a hybrid MPI/OpenMP configuration interaction code for $ab$ $initio$ nuclear structure calculations \cite{SteCON2008}\cite{MarPCS1}\cite{AktCCPE26}. \\

 The properties of $^{7}$Li with JISP16 and NNLO$_{\rm{opt}}$ and $^{7}$Be with JISP16 are given in the next section. Results for $^{7}$Li with JISP16 up to and including the $N_{\rm{max}}=14$ truncation are in agreement with those in Ref. \cite{JPV342}. The values of $^{7}$Li with the NNLO$_{\rm{opt}}$ interaction, as well as the $^7$Li results with JISP16 at $N_{\rm{max}}=16$, are reported here for the first time. The properties we present for $^{7}$Be, the mirror nuclei of $^{7}$Li, are more extensive than those presented in the related references. In order to assess convergence, we present results for even values of the parameter $N_{\rm{max}}$ from 8 to 16, and for $\hbar\Omega$ over a range from 10 MeV to 40 MeV.

\section{Results}

~~~The ground state energy is one of the primary observables in nuclear physics. Our results for the ground state energies of $^{7}$Li with JISP16 and NNLO$_{\rm{opt}}$, and of $^{7}$Be with JISP16, are plotted in Fig. 1 as a function of the HO energy at our selected values of $N_{\rm{max}}$. With increasing $N_{\rm{max}}$, the minima of the U-shaped curves are closer to each other and the shapes exhibit reduced dependence of $\hbar\Omega$. We observe in Fig. 1 that the energies with JISP16 are lower than those with NNLO$_{\rm{opt}}$ for $^{7}$Li at each value of $N_{\rm{max}}$. Moreover, the ground state energies from the JISP16 interaction for both $^{7}$Li and $^{7}$Be show better convergence patterns than that from NNLO$_{\rm{opt}}$. The lowest minima (points closest to convergence) in the three panels are near $\hbar\Omega=20$ MeV. We also indicate the first breakup thresholds in Fig. 1 ($^{3}\rm{H}$+$^{4}\rm{He}$ for $^{7}$Li and $^{3}\rm{He}$+$^{4}\rm{He}$ for $^{7}$Be). Note that we use the experimental energies for the thresholds since both JISP16 and NNLO$_{\rm{opt}}$ give very accurate binding energies for nuclei with $A \leq 4$. The energies with JISP16 in panel 1 and panel 3 are lower than these thresholds. The energy with NNLO$_{\rm{opt}}$ is very close to the threshold where it appears near the tangent of the $N_{\rm{max}}=16$ U-shaped line. We will next see that, upon extrapolation, the ground state energies of all cases in Fig. 1 lie below the lowest thresholds.\\

We now discuss how we extrapolate the results obtained in finite basis spaces to the infinite basis space and produce an extrapolation uncertainty. To obtain the extrapolated ground state energy $E(\infty)$, we use the exponential form\cite{MarPRC79}\cite{JPV342}\cite{BogNPA801}\cite{JPV359}\cite{Shibook}\cite{ForPRC77},
\begin{equation}
E(N_{\rm{max}})=a \rm{Exp}(-c N_{\rm{max}})+E(\infty)
\end{equation}
We follow the method described as "Extrapolation B" in Ref. \cite{MarPRC79}. Under the assumption that the convergence is exponential, we use three successive value of $N_{\rm{max}}$ at each value of $\hbar\Omega$. The numerical uncertainty is the difference between the extrapolated results from two consecutive sets of $N_{\rm{max}}$ values. The extrapolated results using the highest available set of $N_{\rm{max}}$ values (12, 14 and 16) and their numerical uncertainties (depicted as error bars) are shown in Fig. 1. These numerical uncertainties due to extrapolations, also defined in Ref. \cite{MarPRC79}, are minimal and the extrapolations nearly independent of HO energy near $\hbar\Omega =20$ MeV for JISP16 and near 25 MeV for NNLO$_{\rm{opt}}$ indicating favorable convergence estimates at those values of the HO energy.\\

\begin{figure*}
\setlength{\abovecaptionskip}{-0.5cm}
\includegraphics[width=7.4in,height=4.3in]{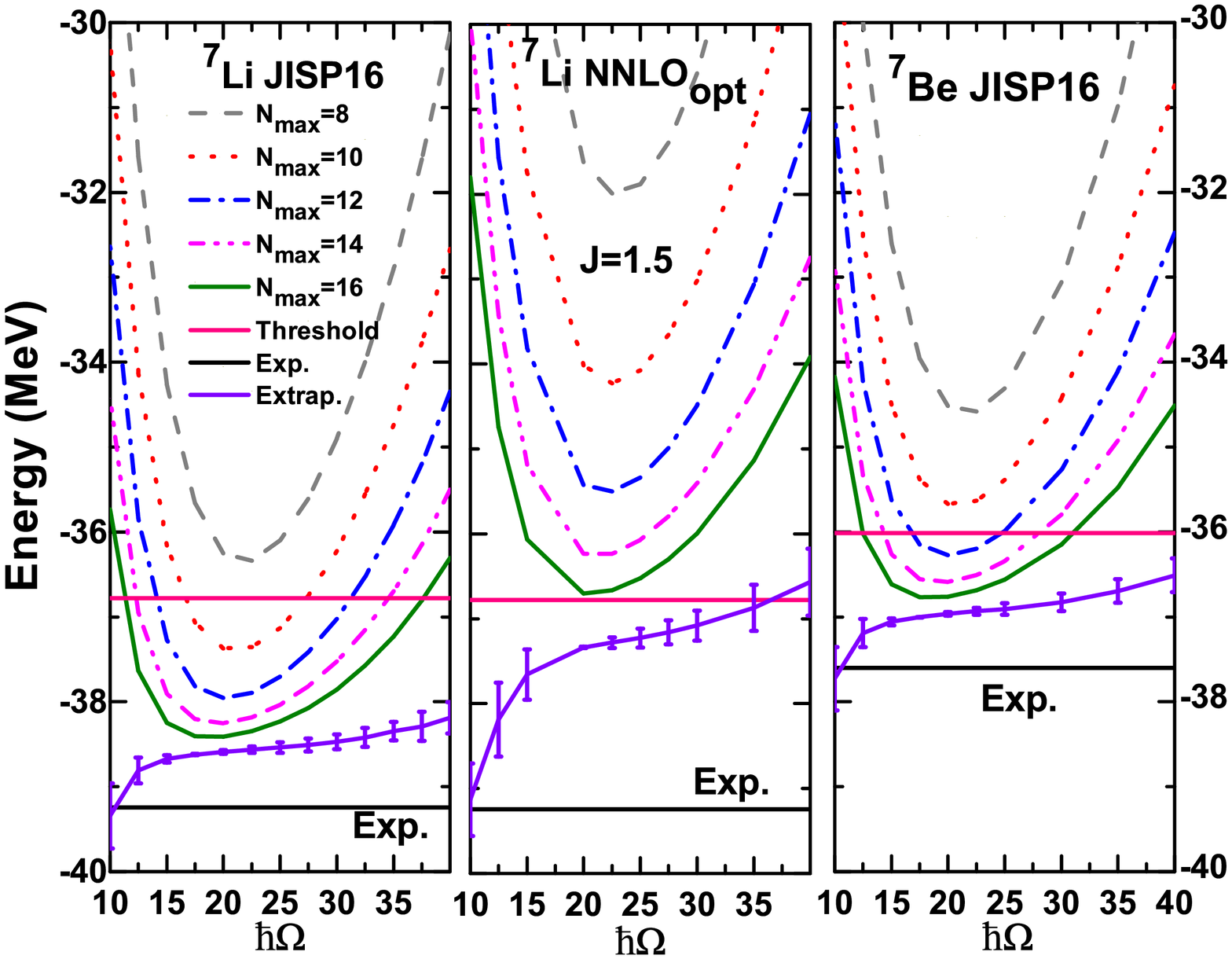}
\caption{\label{fig:wide}(Color online) The energy of the ground state (J=$\frac{3}{2}$) for $^{7}$Be and $^{7}$Li with the JISP16 and NNLO$_{\rm{opt}}$ interactions as a function of HO energy. In this figure and the following figures, for $^{7}$Li and $^{7}$Be, the $N_{\rm{max}}$ value ranges from 8 up to 16. The increment of $N_{\rm{max}}$ is 2. Extrapolated ground state energies are shown in purple with uncertainties depicted as vertical bars.}
\end{figure*}
\begin{figure*}
\setlength{\abovecaptionskip}{-0.5cm}
\includegraphics[width=7.4in]{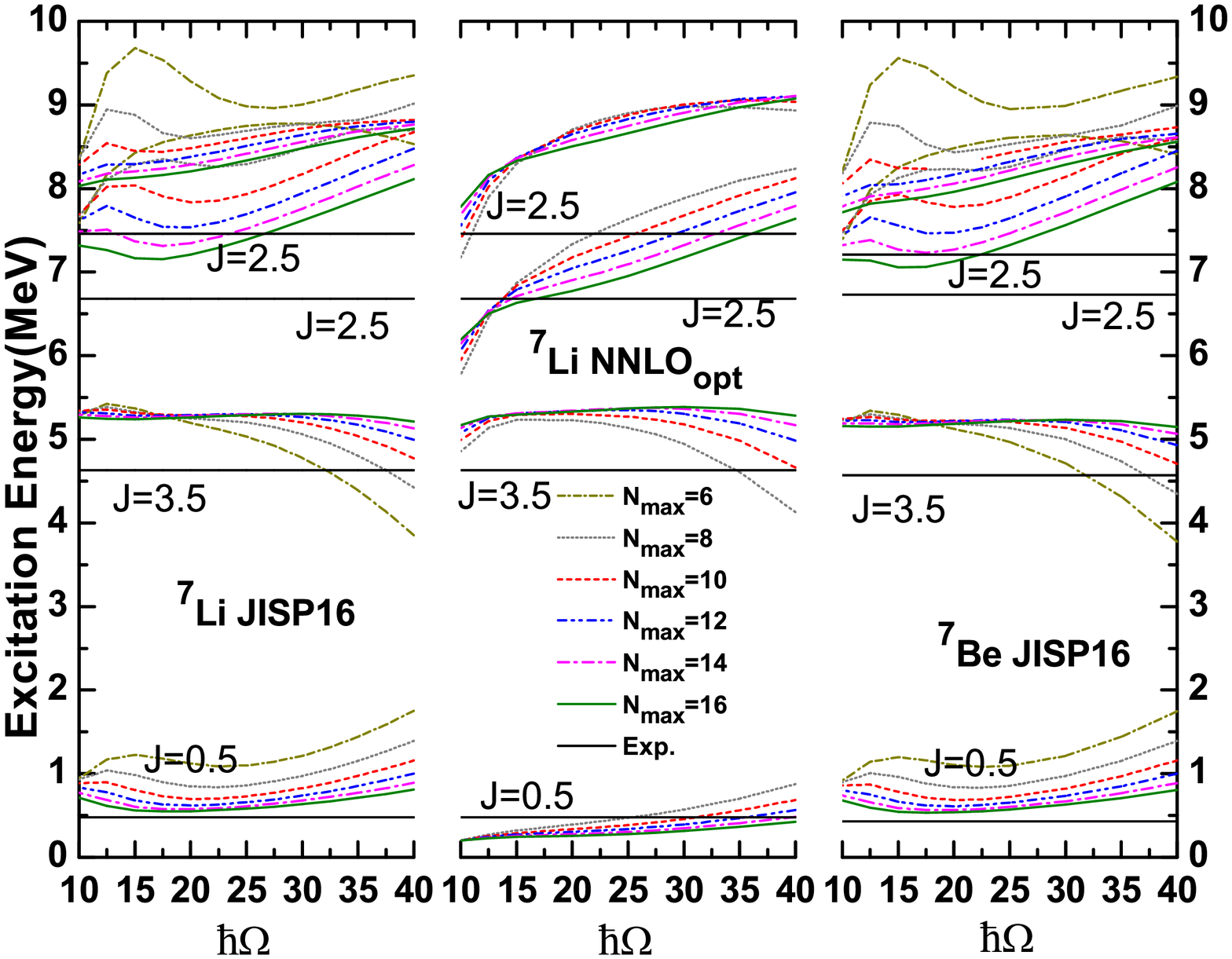}
\caption{\label{fig:wide}(Color online) The energies of the four lowest excited states ($J^{\pi}=\frac{1}{2}^{-}, \frac{7}{2}^{-}, \frac{5}{2}^{-}, \frac{5}{2}^{-},$) with JISP16 and NNLO$_{\rm{opt}}$ for $^{7}$Be and $^{7}$Li as a function of the HO energy. The corresponding experimental values are shown as a horizontal solid line. The Y-axis
indicates the energy gap between the excited and ground state.}
\end{figure*}

We present the energies of the four lowest excited states as a function of the HO energy with a series of $N_{\rm{max}}$ values in Fig. 2 and compare with the experimental excitation energies. The Y-axis denotes the excitation relative to the ground state at the same HO energy and $N_{\rm{max}}$ value. In general, many of the results show reasonable convergence in the $\hbar\Omega$ range of 15 - 25 MeV where the ground state energies are close to their minima. With increasing $\hbar\Omega$, the JISP16 and NNLO$_{\rm{opt}}$ interactions produce similar $\hbar\Omega$-dependence patterns for $^{7}$Be and $^{7}$Li.\\

 We now focus on the upper two excited states which have the same total angular momentum $J=\frac{5}{2}$. In the left and right panels, i.e. for $^7$Li and $^7$Be respectively with JISP16, the two states are nearly degenerate. However, with the NNLO$_{\rm{opt}}$ interaction for $^7$Li (central panel), there is about a 1 MeV larger energy gap between the two states. In addition, both the $J=\frac{5}{2}$ curves for NNLO$_{\rm{opt}}$ increase monotonically with increasing $\hbar\Omega$ while the corresponding JISP16 curves display peaks and dips in the lower $\hbar\Omega$ range which become less pronounced with increasing $N_{\rm{max}}$.
 These non-monotonic features in the case of JISP16 may be attributed, at least in part, to the mixing of these two states as
 we discuss below.\\

\begin{figure*}
\setlength{\abovecaptionskip}{-0.5cm}
\includegraphics[width=7.4in,height=4.3in]{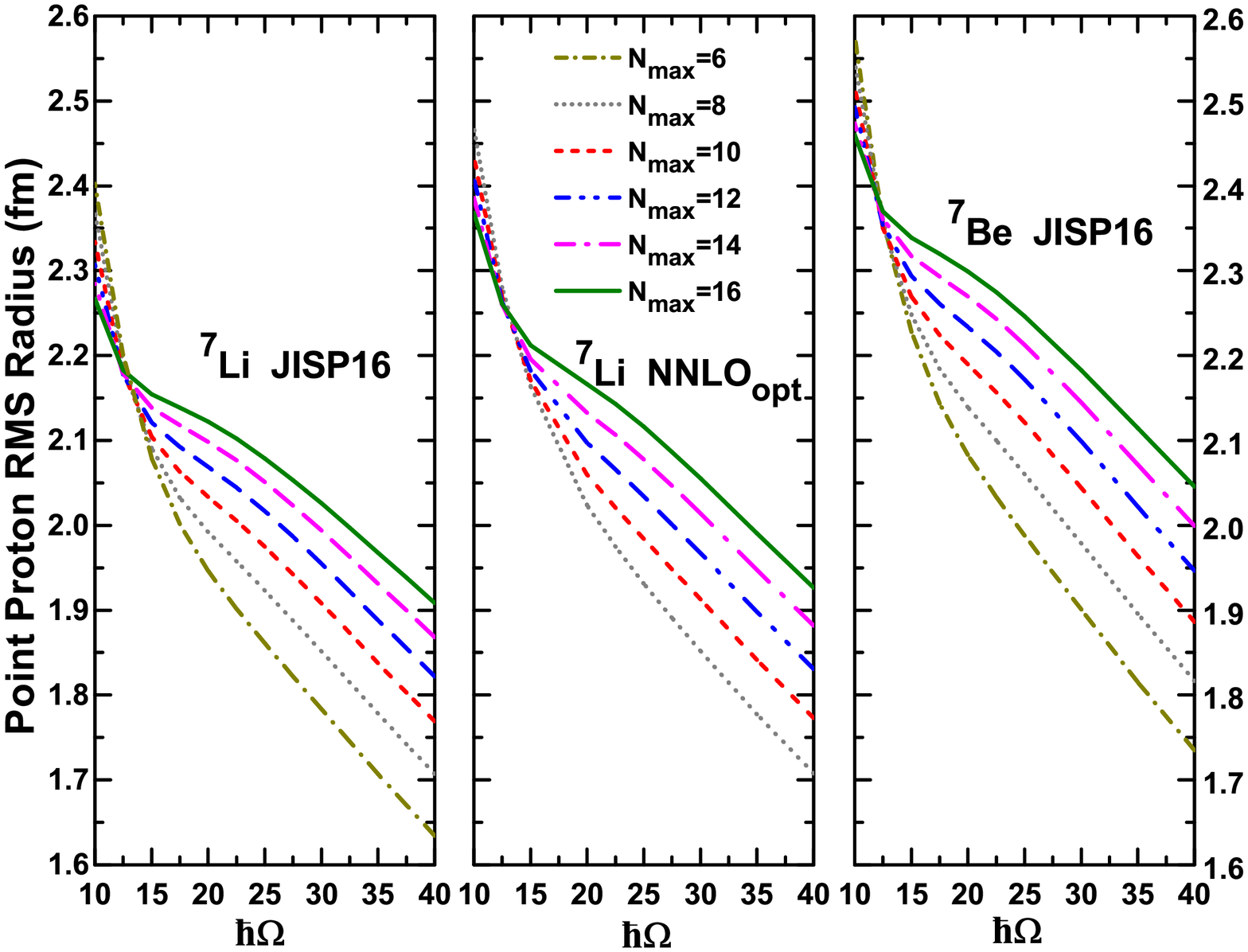}
\caption{\label{fig:wide}(Color online)Point proton root-mean-square (rms) radius (in fm) of the ground state as a function of HO energy with JISP16 and NNLO$_{\rm{opt}}$ for a sequence of $N_{max}$ values. The left and right graphs are with JISP16, and the middle graph is with NNLO$_{\rm{opt}}$.}
\end{figure*}

Let us now turn our attention to observables other than the energies. The calculated point proton root-mean-square (rms) radii are presented for the ground states of $^{7}$Li and $^{7}$Be as a function of HO energy for a range of $N_{\rm{max}}$ values in Fig. 3. We first note that these rms radii are approximately independent of $N_{\rm{max}}$ near $\hbar\Omega=12$ MeV. This feature of the rms radii, sometimes referred to as defining the "interaction" or "crossover" point, has been noted in Refs. \cite{BogNPA801}\cite{NogPRC73}\cite{CapPRC86} and is sometimes taken as the estimate of the converged rms radius. We adopt the practice of quoting the crossover point of the rms radii in the current work since the robust extrapolations of rms radii will require additional developments and/or larger basis spaces. Similarly, for other observables quoted below, we cite only an estimate based on a visual inspection of the convergence pattern without an associated uncertainty. Over time, we anticipate that further theoretical developments will provide reliable extrapolation tools that yield refined converged estimates from our results.\\

 With increasing $\hbar\Omega$, the patterns of the rms radii are similar for $^{7}$Be and $^{7}$Li for the two different interactions. The radii of $^{7}$Li and $^{7}$Be decrease monotonically with increasing $\hbar\Omega$ and show a weak trend towards convergence as is frequently found for this long-range observable calculated in a HO basis. The value of the rms radius is a little larger with NNLO$_{\rm{opt}}$ than that with JISP16 for $^{7}$Li at the same $\hbar\Omega$ and $N_{\rm{max}}$ values, which may simply be a reflection of the lower binding energy of NNLO$_{\rm{opt}}$ relative to JISP16.\\

The magnetic dipole moments in units of $\mu_N$ are depicted in Fig. 4 as a function of the HO energy for $^{7}$Li and $^{7}$Be with the JISP16 and NNLO$_{\rm{opt}}$ interactions. It is easy to see that the magnetic dipole moment for $^{7}$Be is negative and those for $^{7}$Li are about $3 \mu_{N}$ with the two different interactions. With increasing $N_{\rm{max}}$, the magnetic moments tend to decrease slightly in magnitude but converge to within about 1-2$\%$. The converged value of $^{7}$Be appears to be a little below $-1.08\mu_N$. The converged magnetic dipole moment of $^{7}$Li with NNLO$_{\rm{opt}}$ is close to $2.97\mu_N$ while the result with JISP16 is a little less than $2.96\mu_N$.\\

Our results for the ground state quadrupole moments, another important electromagnetic observable, are shown in Fig. 5. The patterns for the quadrupole moment results reflect the patterns seen above for the rms radii since they are closely related observables. In particular, we again observe a very weak convergence pattern with increasing $N_{\rm{max}}$. One may visually estimate that the converged results will be in the region of -3.5 $\rm{e}\cdot \rm{fm}^2$ for $^7$Li and -5.5 $\rm{e}\cdot \rm{fm}^2$ for $^7$Be.\\

Next, we present the results of two B(E2) transitions to the ground state ($J^{\pi}=\frac{3}{2}^{-}$) in Fig. 6 where the transition from the $J^{\pi}=\frac{1}{2}^{-}$ state is represented by the top three panels and from the $J^{\pi}=\frac{7}{2}^{-}$ state by the bottom three panels. The left panels and the middle panels are for $^{7}$Li with JISP16 and NNLO$_{\rm{opt}}$ respectively. The right panels are for $^{7}$Be with JISP16. As may be expected, the patterns of these B(E2) transitions are also similar to those of point proton rms radii in Fig. 3. We again observe a very weak tendency for convergence reflecting the major role of the radius operator in the B(E2). Comparing both $J^{\pi}=\frac{1}{2}^{-}$ and $\frac{7}{2}^{-}$ $\rightarrow$ ground state (signified by "g.s." in the label) B(E2) transitions for $^{7}$Li (with either the JISP16 and NNLO$_{\rm{opt}}$ interactions), the results for the latter transition 
are about half those of the former transition. This pattern persists in the $^{7}$Be transitions also seen in Fig. 3.\\

The excitation energies, quadrupole moments and B(E2)'s that we have presented and discussed for $^{7}$Be are key observables that play a critical role in the identification and characterization of its yrast rotational band \cite{Caprio:2013yp,Maris:2014jha,Caprio:2015iqa,Caprio:2015jga}.  Indeed, their systematics along with the systematics of the other members of the same band obtained in NCFC calculations with JISP16 are essential to demonstrating emergence of collective rotational motion in $^{7}$Be.

We present three B(M1) transitions as a function of the HO energy in Fig. 7 for $^{7}$Li and $^{7}$Be with a sequence of $N_{\rm{max}}$ values. The three top graphs display the B(M1) transitions from the $J^{\pi}=\frac{1}{2}^{-}$ state to the  $J^{\pi}=\frac{3}{2}^{-}$ ground state. The three middle graphs and the three bottom graphs are from the 
$J^{\pi}=\frac{5}{2}^{-}_{1}$ state and the $J^{\pi}=\frac{5}{2}^{-}_{2}$ state to the ground state, respectively. The subscript $1$ ($2$) on the $\frac{5}{2}^{-}$ signifies the lower (upper) of the two states with $J^{\pi}=\frac{5}{2}^{-}$.\\

It is noteworthy that the top three graphs have the same convergence pattern for $^{7}$Li and $^{7}$Be with the JISP16 and NNLO$_{\rm{opt}}$ interactions. Considering the greatly expanded scales used for these B(M1) results, one observes that good convergence is actually attained in all cases shown in Fig. 7.  In particular the convergence at the highest $N_{\rm{max}}$ shown is good over a fairly large range in $\hbar\Omega$ from about 15 MeV to about 35 MeV.  These B(M1)'s as well as the magnetic dipole moments continue to be among the best converged of the electromagnetic observables in NCFC calculations.\\

Features suggestive of the mixing of the two $\frac{5}{2}^{-}$ states, which were discussed above in connection with the behaviors of the excitation energies, are also apparent in the B(M1) transitions of Fig. 7. The low and the high 
$\hbar\Omega$ regions of the $^7$Li transitions from these $\frac{5}{2}^{-}$ states with the JISP16 interaction, for example, appear to support the discussions of mixing that were stimulated by the results for the excitation energies.
This mixing is again seen to decrease with increasing $N_{max}$.
In addition, one may now interpret the results for the B(M1)s from the two $\frac{5}{2}^{-}$ states in $^7$Be as suggesting mixing that also decreases with increasing $N_{max}$.

In order to better examine the nuclear structure and the relationship between the different states, we present the total magnetic moment and the contributions to the total angular momentum from the orbital motions of the proton and neutron as well as the contributions from intrinsic spin in Fig. 8.  We follow the procedures presented in Ref. \cite{JPV359} and define these contributions though matrix elements of the projections of these individual contributions on the state's total angular momentum, i.e. by matrix elements of the terms on the right-hand side of the following equation:\\
\begin{equation}
J=\frac{1}{J+1}(<\vec{J}\cdot \vec{L}_p>+<\vec{J}\cdot \vec{L}_n>+<\vec{J}\cdot \vec{S}_p>+<\vec{J}\cdot \vec{S}_n>).
\end{equation}
 From top to bottom, the three graphs are for $^{7}$Li with JISP16, $^{7}$Li with NNLO$_{\rm{opt}}$ and $^{7}$Be with JISP16 respectively. One may compare the top panel in Fig. 8 for $^{7}$Li with JISP16 to corresponding results in Ref. \cite{JPV359} where the results were shown through $N_{\rm{max}}=14$. From the left to the right in each panel, the individual frames are for the states $J^{\pi}=\frac{3}{2}^{-}, \frac{1}{2}^{-}, \frac{7}{2}^{-}, \frac{5}{2}^{-}_{1}$ and $\frac{5}{2}^{-}_{2}$ successively. In the frames for the first three states, we see that the four components are well converged as they become nearly independent of the values of $\hbar\Omega$ and $N_{\rm{max}}$ with increasing $N_{\rm{max}}$. \\

 In the two panels with the JISP16 interaction results in Fig. 8, the two states with $J=\frac{5}{2}$ exhibit trends suggesting there is crossing and mixing as a function of $\hbar\Omega$ at lower values of $N_{\rm max}$. We have elected to display a larger range of results in Fig. 8 for these two states than presented in Ref. \cite{JPV359} in order to map out this level crossing as a function of $N_{\rm{max}}$ and $\hbar\Omega$.  Clearly with increasing $N_{\rm{max}}$, the level crossing as a function of $\hbar\Omega$ is disappearing in 
 the $\hbar\Omega$ range of 10 to 40 MeV.\\

We also include the magnetic moment in Fig. 8 for each state (purple symbols and lines) which is a sum over the weighted contributions from the other results depicted. Thus, the magnetic moment shows a similar convergence pattern to those of its individual contributions.\\

  Let us examine the results for the $\frac{7}{2}^-$ state in Fig. 8 in some detail.
In all cases, the intrinsic spin (green) provides almost no contribution to the total angular momentum. For $^7$Li the neutron orbital motion (blue) dominates while for $^7$Be the proton orbital motion (red) dominates as may be expected on the basis of isopsin symmetry (an approximate symmetry since we include Coulomb and NNLO$_{\rm{opt}}$ is charge-dependent). Due to the relative weightings of these contributions, the magnetic moment shows an apparent slower trend towards convergence.  For example, if we take the case of $^7$Li with NNLO$_{\rm{opt}}$, we see that the $\frac{7}{2}^-$ state has a negative contribution from the neutron intrinsic spin that, due to the large weighting of the neutron intrinsic magnetic moment, provides a slower convergence pattern than observed for the $\frac{3}{2}^-$ and $\frac{1}{2}^-$ states. The source of this weaker convergence is the weaker convergence of the neutron spin contribution (green) in the $\frac{7}{2}^-$ state compared to its contribution in the two lower-lying states - though it may be challenging to see this detail in the central frame of the middle panel in Fig. 8.\\

 Finally, the properties of $^{7}$Li with the JISP16 and NNLO$_{\rm{opt}}$ interactions and $^{7}$Be with JISP16 are listed in Table.1. The results are compared to the experimental values as well as those from AV18+IL7\cite{Shibook}\cite{PasPRC87} and chiral NN+NNN\cite{MarPRC87}. For $^{7}$Li, the results with different interactions are listed in columns 2-6, and those for $^{7}$Be are in columns 7-9. In the present work, we do not quote the uncertainty which has been explained and discussed in related references. However, the uncertainties estimated in Ref. \cite{MarPRC87} remain as valid estimates for our current uncertainties (see, for example, the extrapolation uncertainties depicted in Fig. 1) so that we can assert the following: the theoretical results with JISP16 and NNLO$_{\rm{opt}}$ interaction are in reasonable agreement with the experimental results. Nevertheless, the results with AV18+IL7 appear to be more consistent with experiment. The values in the fourth column for NNLO$_{\rm{opt}}$ show somewhat larger differences with experiment than those with JISP16 and AV18+IL7 possibly due, primarily, to slower convergence of observables in $^7$Li with this interaction.

\section{Summary}

We have calculated the properties of $^{7}$Li and $^{7}$Be with the JISP16 and NNLO$_{\rm{opt}}$ interactions in the no-core full configuration (NCFC) approach. We present  results with the many-body truncation parameter $N_{\rm{max}}$ up through 16 and for natural parity ($N_{\rm{max}}$ is even). We obtained the energies of the ground and excited states, point proton rms radii, magnetic and quadrupole moments, E2 and M1 transitions. To our knowledge, this is the first time that properties of $^{7}$Li with NNLO$_{\rm{opt}}$ are presented. The theoretical results and experimental data are compared in Table. 1. Taking into consideration the attained degree of convergence, we find the results with JISP16 or AV18+IL7 are in better agreement with the experimental data than those with NNLO$_{\rm{opt}}$. In order to aid in the diagnostics of individual states and to better understand the generally good convergence of magnetic moments, we decompose the magnetic moment contributions into proton and neutron orbital and spin components.  This aids us, for example, to observe level crossing in the $J = \frac{5}{2}$ states found with the JISP16 interaction in both $^7$Li and $^7$Be.  Interestingly, we did not observe this level crossing in $^7$Li with NNLO$_{\rm{opt}}$.

\section{Acknowledgments}

We acknowledge useful discussions with Andrey Shirokov and Mark Caprio. This work was supported by the National Natural Science Foundation of China (grant no. 11205004),
by the US Department of Energy under Grants No. DESC0008485 (SciDAC/NUCLEI) and
No. DE-FG02-87ER40371, by the US National Science Foundation under Grant No. 0904782.
Computational resources were provided by the National Energy Research Supercomputer
Center (NERSC), which is supported by the Office of Science of the U.S. Department
of Energy under Contract No. DE-AC02-05CH11231. Computational resources were also provided by the Argonne Leadership Computing Facility (ALCF) (US DOE Contracts No. DE-AC02-05CH11231 and DE-AC02-06CH11357) and under an INCITE award (US DOE Office of Advanced Scientific Computing Research).


\begin{figure*}
\setlength{\abovecaptionskip}{-0.5cm}
\includegraphics[width=7.4in,height=4.2in]{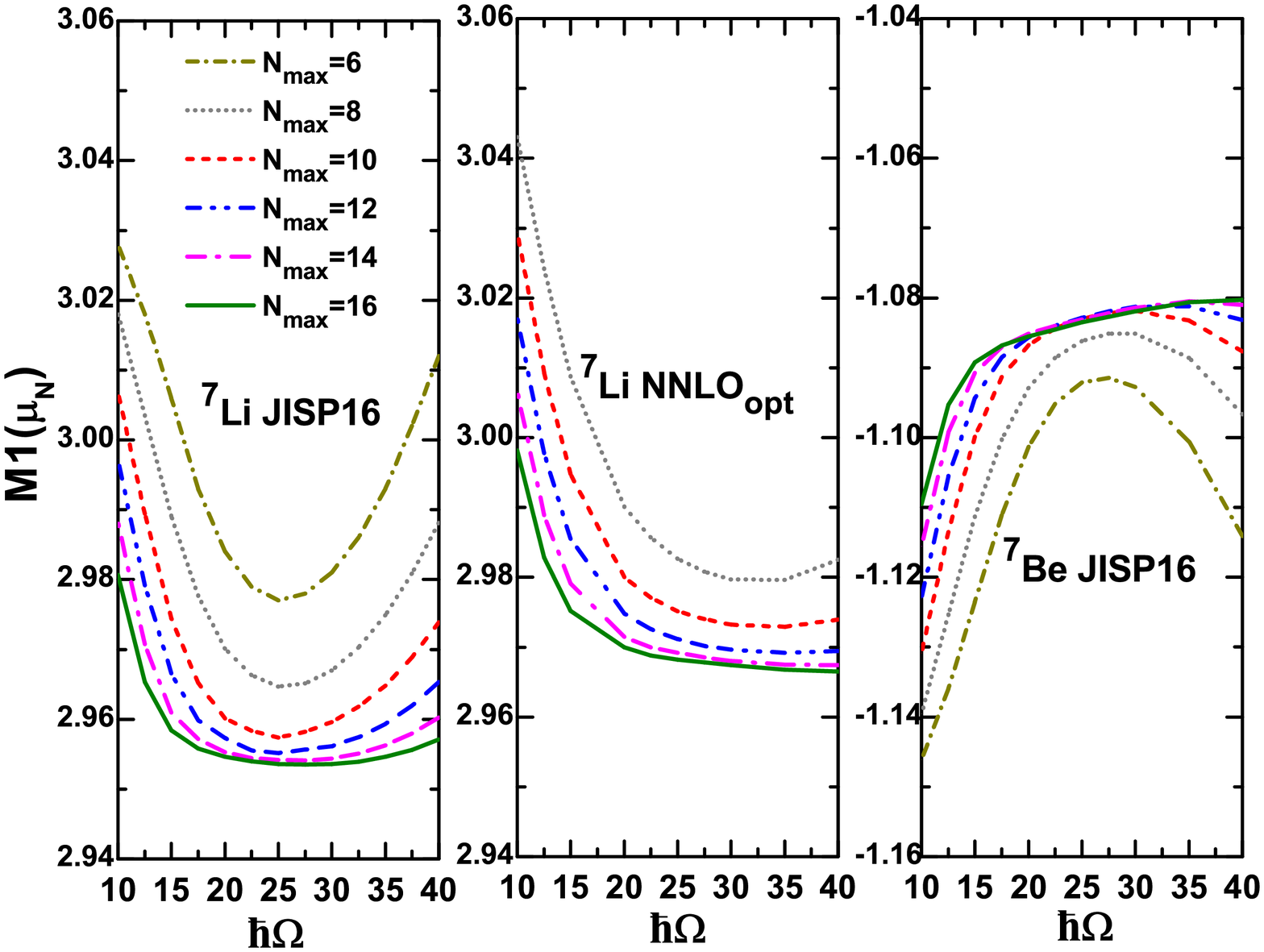}
\caption{\label{fig:wide}(Color online)Magnetic dipole moment of the ground state shown as a function of the HO energy for $^{7}$Be and $^{7}$Li with the JISP16 and NNLO$_{\rm{opt}}$ interactions.}
\end{figure*}

\begin{figure*}
\setlength{\abovecaptionskip}{-0.3cm}
\includegraphics[width=7.4in,height=4.5in]{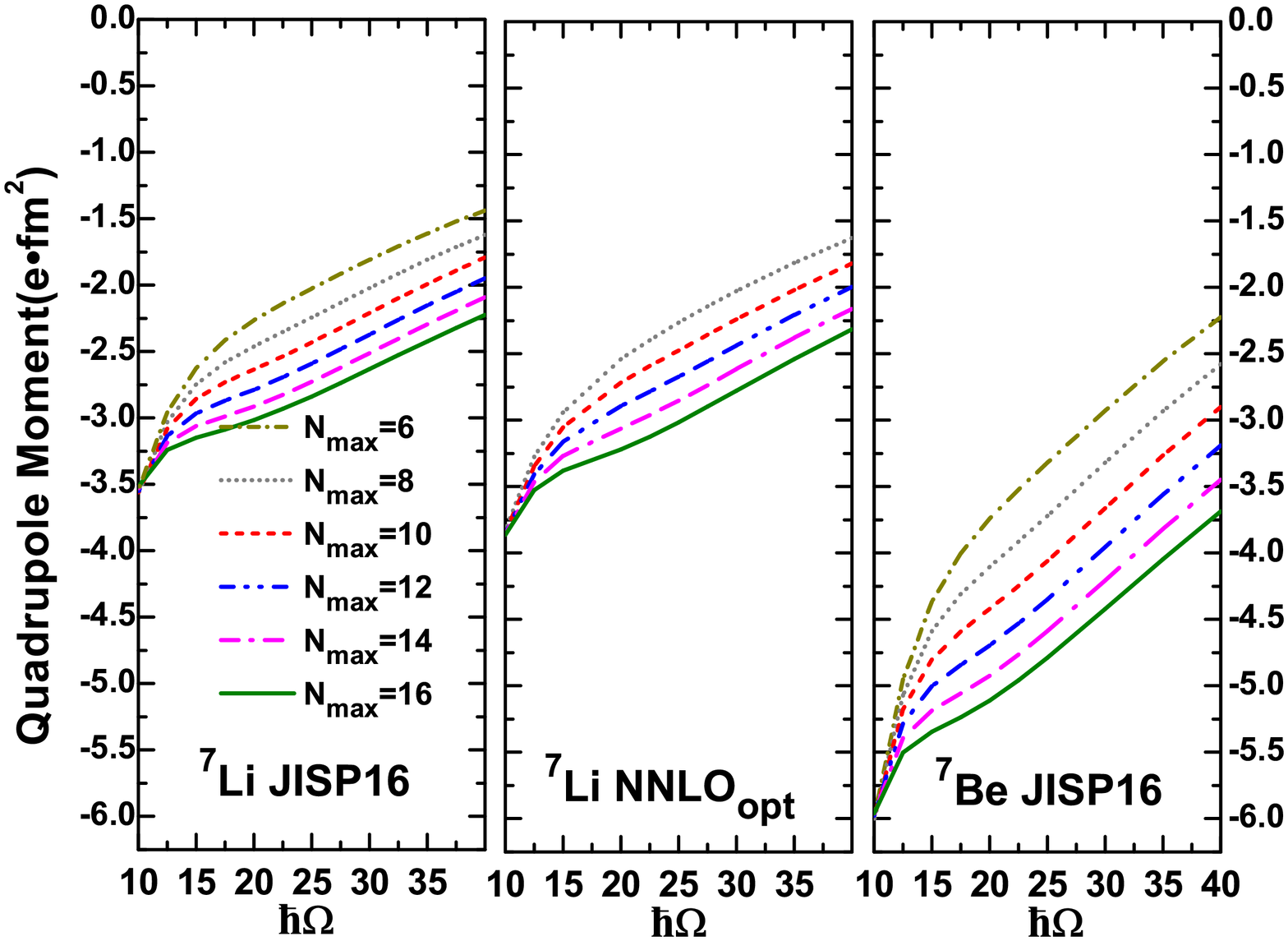}
\caption{\label{fig:wide}(Color online) Quadrupole moment of the ground state as a function of the HO energy with a sequence of $N_{\rm{max}}$ values for $^{7}$Be and $^{7}$Li.}
\end{figure*}

\begin{figure*}
\setlength{\abovecaptionskip}{-1cm}
\includegraphics[width=7.4in]{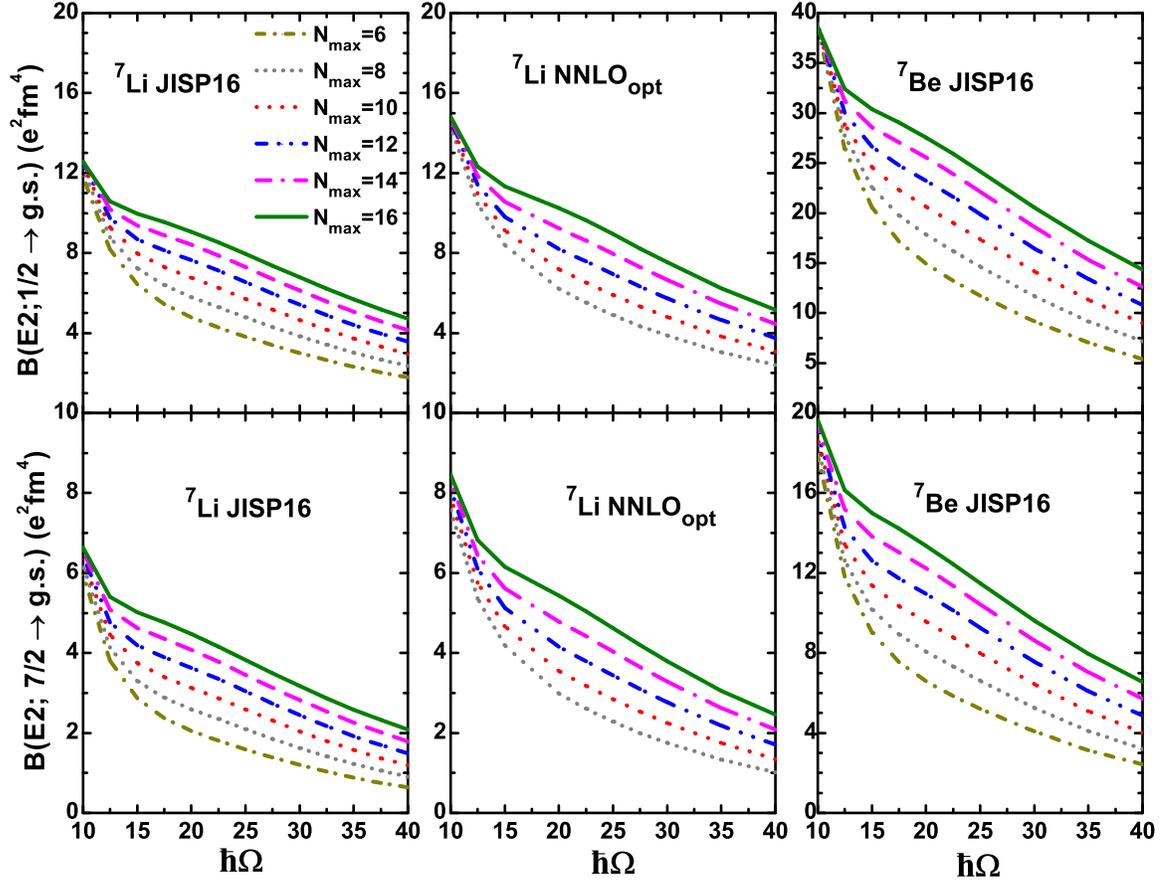}
\caption{\label{fig:wide}(Color online) B(E2) to the ground state ($J^{\pi}=\frac{3}{2}^{-}$) from the states $J^{\pi}=\frac{1}{2}^{-}$ (top) and $J^{\pi}=\frac{7}{2}^{-}$ (bottom) as a function of the HO energy with a sequence of $N_{\rm{max}}$ values. The abbreviation "g.s." signifies the ground state.}
\end{figure*}

\begin{figure*}
\setlength{\abovecaptionskip}{-1cm}
\includegraphics[width=7.4in]{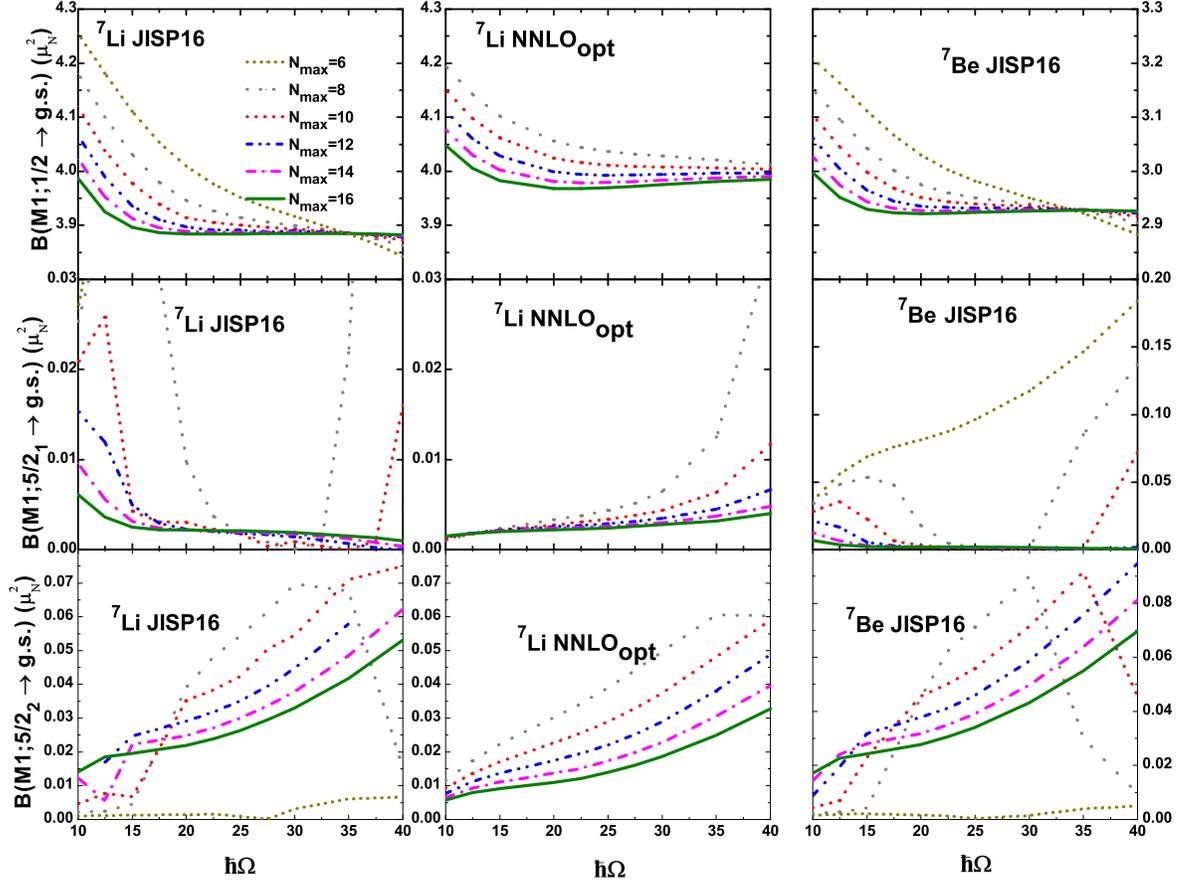}
\caption{\label{fig:wide}(Color online) B(M1) to the ground state ($J^{\pi}=\frac{3}{2}^{-}$) from the states ($J^{\pi}=\frac{1}{2}^{-}$) (top),  ($J^{\pi}=\frac{5}{2}^{-}_{1}$) (middle) and ($J^{\pi}=\frac{5}{2}^{-}_{2}$) (bottom) as a function of the HO energy for $^{7}$Li and $^{7}$Be with JISP16 and NNLO$_{\rm{opt}}$. Note the expanded scales used to reveal similarities and differences in the details.}
\end{figure*}

\begin{figure*}
\setlength{\abovecaptionskip}{-1cm}
\includegraphics[width=7.4in,height=4in]{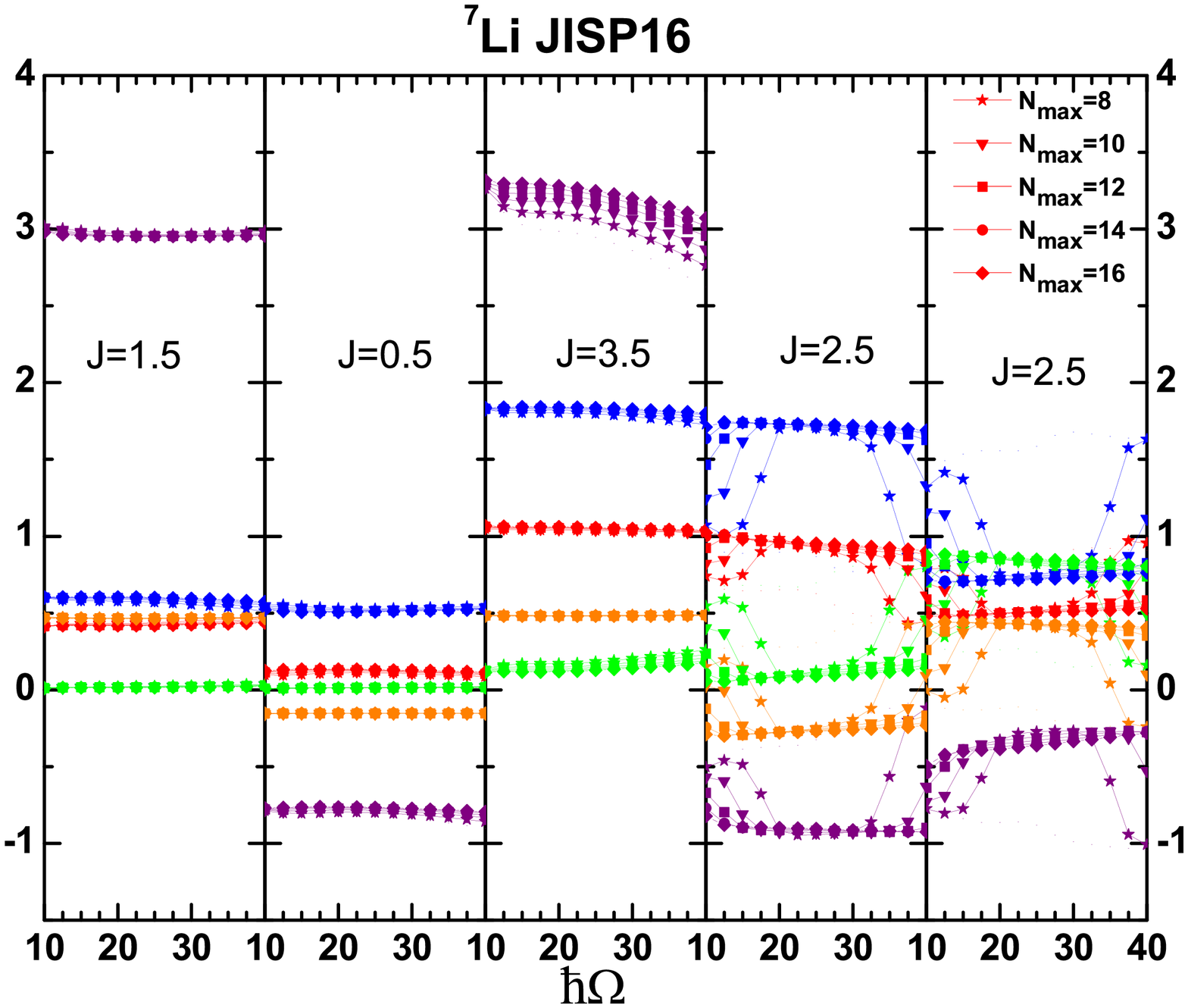}
\end{figure*}
\begin{figure*}
\setlength{\abovecaptionskip}{-1cm}
\includegraphics[width=7.4in,height=4in]{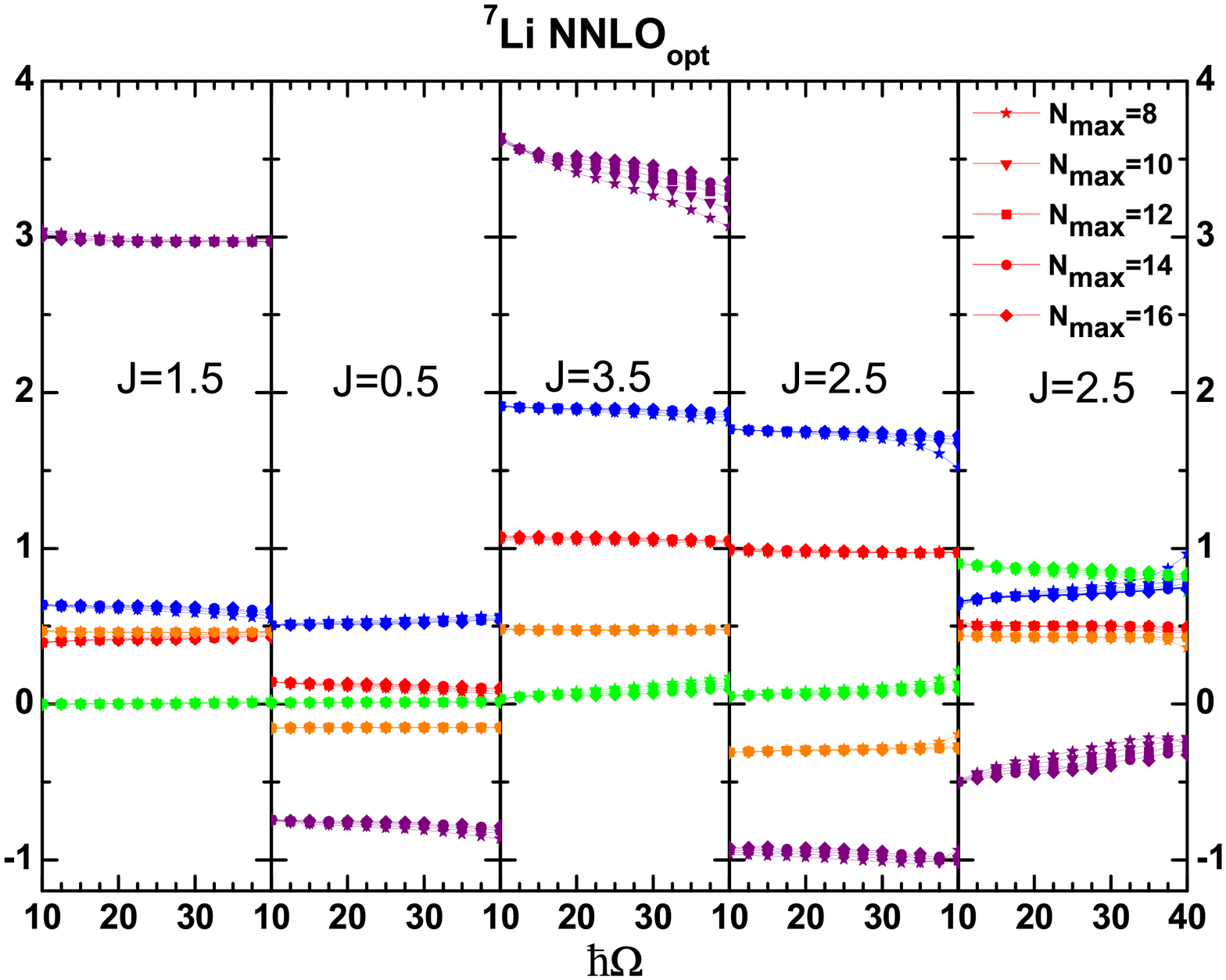}
\end{figure*}
\begin{figure*}
\setlength{\abovecaptionskip}{-1cm}
\includegraphics[width=7.4in]{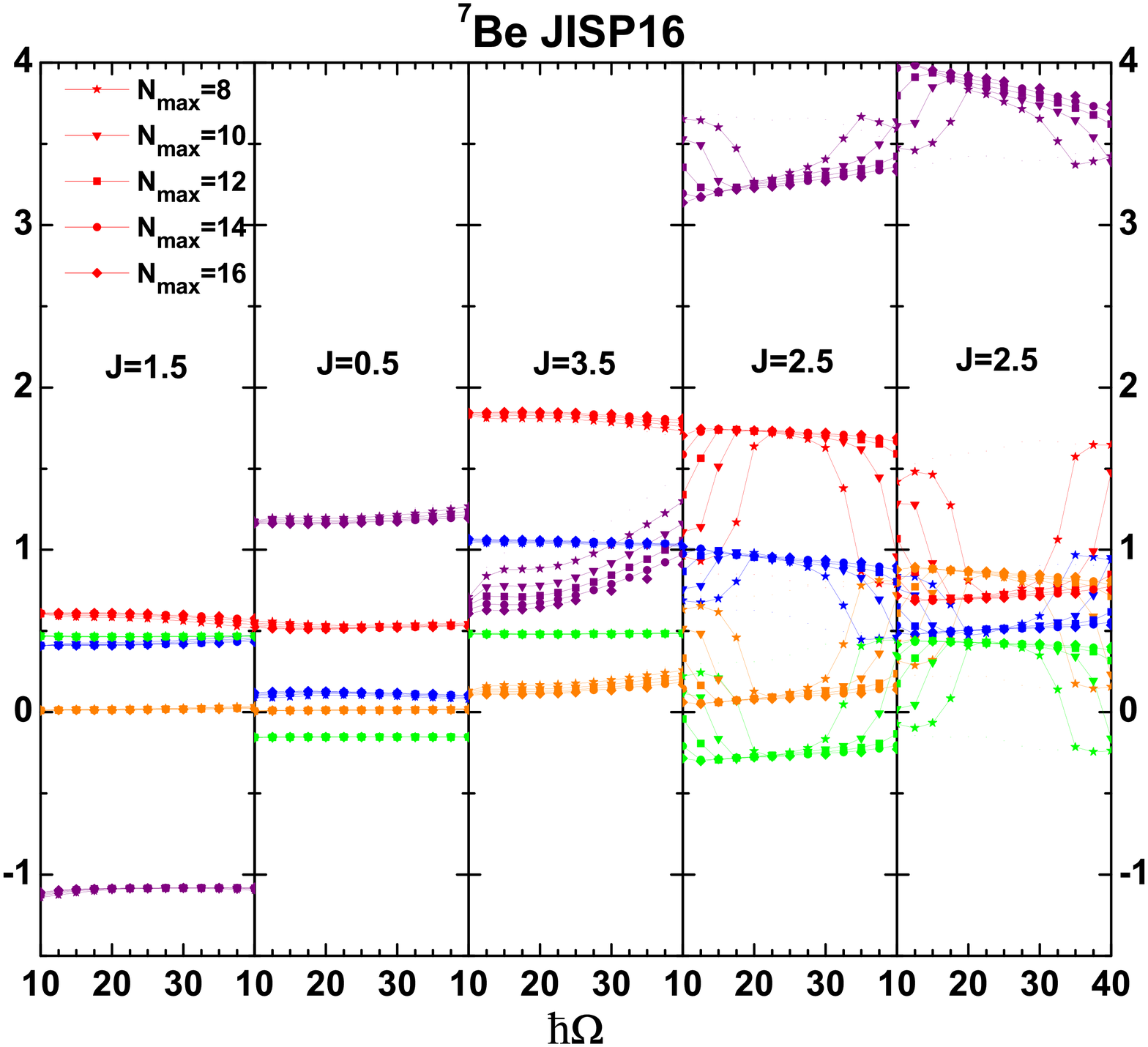}
\caption{\label{fig:wide}(Color online)The three multiframe panels from top to bottom describe the spin decomposition and total magnetic moment for $^{7}$Li with JISP16, $^{7}$Li with NNLO$_{\rm{opt}}$ and $^{7}$Be with JISP16 respectively. They are presented as a function of the HO energy with a sequence of $N_{\rm{max}}$ values. We show the contributions from the proton orbital motion (red), neutron orbital motion (blue), proton intrinsic spin (orange), and neutron intrinsic spin (green). The purple symbols with connecting lines present the total magnetic moment in units of $\mu_N$.}
\end{figure*}

\begin{table*}[width=7in]
\renewcommand\arraystretch{1.5}
\caption{\label{tab:table3}Experimental data and the corresponding theoretical results for $^{7}$Li and $^{7}$Be using the chiral NN+NNN, AV18+IL7, JISP16 and NNLO$_{\rm{opt}}$ interactions. Energies are in MeV, point proton rms radii are in femtometers, quadrupole moments Q are in $\rm{e}\cdot \rm{fm}^2$, $\mu$ are in $\mu_N$, B(E2) are in $\rm{e}^{2}\rm{fm}^{4}$, and B(M1) are in $\mu_{N}^{2}$. Columns 2-6 are for $^{7}$Li and Columns 7-9 are for $^{7}$Be. Our calculations for $^{7}$Li are with JISP16 and NNLO$_{\rm{opt}}$, $^{7}$Be calculations are with JISP16 and all are at $N_{\rm{max}}=16$. The results of point proton rms radii, quadrupole moments and B(E2)'s are at $\hbar\Omega =10$ MeV, and the other results are at $\hbar\Omega =20$ MeV all at $N_{max}=16$. This $N_{max}$ value and these $\hbar\Omega$ values are adopted as we estimate this is where the cited results are closest to convergence. Our extrapolated ground state energies ($E_{gs}^{extrap}$) are taken
from the results in Fig. 1 where the extrapolation uncertainty is estimated
following the methods of Ref. \cite{JPV359}. The uncertainty is given in parenthesis
for the last significant figure quoted. We also quote uncertainties from the
cited references for other observables. The experimental data are from Ref. \cite{ENE} for energies, from Ref. \cite{JPV342}\cite{PasPRC87} for the other observables. The chiral NN+NNN results come from Ref. \cite{MarPRC87}. (OLS is the abbreviation of Okubo-Lee-Suzuki, the renormalization method employed). The results for AV18+IL7 are from Refs. \cite{PasPRC87}.}

\begin{tabular}{ccccccccccc}
\hline\hline
 &\multicolumn{6}{c}{$^{7}$Li}&\multicolumn{4}{c}{$^{7}$Be}\\
Observable& ~~~~~ ~&Exp.&JISP16&NNLO$_{\rm{opt}}$&AV18/IL7&Chiral NN+NNN(OLS)& ~~~~~ ~&Exp.&JISP16&AV18/IL7\\ \hline
 $E_{gs}$($\frac{3}{2}^{-}$)& ~~~~ ~&39.245&38.412&36.703&39.0(1)&38.60(44)& ~~~~~&37.601&36.763&37.4(1) \\
 $E^{extrap}_{gs}$&~~&~~&38.59(7)&37.3(2)&~~~&~~~&~~&~~&36.97(8)&~~~\\
 $<r^{2}_{p}>^{1/2}$& ~~~~ ~&2.31&2.267&2.368&2.28&2.11&~~~~~ ~&2.51&2.46&2.47\\
$E_x$($\frac{1}{2}^{-}$)& ~~~~ ~&0.478&0.551&0.258&0.1(1)&0.382(69;24)& ~~~~~ ~&0.429&0.538&0.1(1)\\
$E_x$($\frac{7}{2}^{-}$)& ~~~~ ~&4.630&5.262&5.33&4.9(1)&5.20(22;12)&~~~~~ ~&4.57&5.184&-\\
$E^{1}_x$($\frac{5}{2}^{-}$)& ~~~~ ~&6.680&7.209&6.776&6.5(1)&7.50(16;23)& ~~~~~ ~&6.73&7.126&-\\
$E^{2}_x$($\frac{5}{2}^{-}$)& ~~~~ ~&7.460&8.207&8.502&7.7(2)&8.31(01;17)& ~~~~~ ~&7.21&7.959&-\\
$Q_{g.s.}$& ~~~~ ~&-4.00&-3.517&-3.864&-4.0(1)&-2.75&~~~~~ ~&-&-5.961&-6.7(1)\\
$\mu$ & ~~~~ ~&3.256&2.955&2.97&3.24(1)&2.993& ~~~~~ ~&-1.398&-1.086&-1.42(1)\\
B(E2;$\frac{1}{2}^{-}$)& ~~~~ ~&15.7&12.572&14.822&-&7.30& ~~~~~ ~&-&38.603&-\\
B(E2;$\frac{7}{2}^{-}$)& ~~~~ ~&3.4&6.635&8.457&-&3.4& ~~~~~ ~&-&19.637&-\\
B(M1;$\frac{1}{2}^{-}$)& ~~~~ ~&4.92&3.884&3.968&3.13(2)&4.068& ~~~~~ ~&-&2.922&2.72(2)\\
B(M1;$\frac{5}{2}_{1}^{-}$)& ~~~~ ~&-&0.0022&0.0022&-&0.004& ~~~~~ ~&-&0.0019&-\\
\hline
\end{tabular}
\end{table*}

\end{document}